# Mobile Services and Network Proximity

Dmitry Namiot
Lomonosov Moscow State University
Faculty of Computational Mathematics and Cybernetics
Moscow, Russia
dnamiot@gmail.com

Manfred Sneps-Sneppe
Ventspils University College
Ventspils International Radioastronomy Centre
Ventspils, Latvia
manfreds.sneps@gmail.com

*Abstract*— **This paper discusses several practical use cases for deploying network proximity in mobile services. Our research presents here mobile services oriented for either discovering new data for mobile subscribers or for delivering some customized information to them. All applications share the same approach and use the common platform, based on the Wi-Fi proximity. The typical deployment areas for our approach are context-aware services and ubiquitous computing applications. Our own examples include proximity marketing as the model use case.**

*Keywords - context-aware;proximity;Wi-Fi;ubiquitous computing;productions*

## I. INTRODUCTION

Classically, term 'context-aware' [1] describes context as location, identities of nearby people and objects, and changes to those objects. Indeed, most of the authors define context awareness as a complementary element to location awareness. Location serves as a determinant for the main processes and context adds more flexibility with mobile computing and smart communicators [2].

In the same time, there are many practical use cases, where the concept of location can be replaced by that of proximity. At the first hand, this applies to use cases where the detection for exact location is difficult, even impossible or not economically viable [2]. Another reason for such replacement is related to the privacy. For example, we can think here about a privacy-aware proximity detection service determines if two mobile users are close to each other without requiring them to disclose their exact locations. Proximity can be used as a main discovery element for context-aware browser [3]. In this concept, any existing or even especially created Wi-Fi hot spot or Bluetooth node could be used as a presence sensor that can play a role of presence trigger. This trigger can open access to some content, discover existing content as well as cluster the nearby mobile users.

A definition for network proximity could be very transparent. It is a measure of how mobile nodes are close or far away from the elements of network infrastructure. In this definition, the elements of network infrastructure (e.g., Wi-Fi access points) play a role of landmarks. For some tasks we need to measure how our mobile clients are close each other, using the network infrastructure information only. The classical model uses Bayesian reasoning and a hidden Markov model (HMM) [4]. Other models took into account not only signal strengths, but also the probability of seeing an access point from a given location. But the biggest practical problem for such models is the mandatory manual calibration phase. Classical Wi-Fi based positioning system works in two phases. Phase 1 is the offline training phase or calibration. In phase 1 a human operator measures the received signal strength indicator (RSSI) from different access points (APs) at selected positions in the environment. This phase creates a radio map that stores the RSSI values of access points at different fixed points. And only phase 2 is the real location estimation. And of course, any changes in the network (network's infrastructure) cause recalibration process. Another problem that could not be solved with calibration (Wi-Fi fingerprints collection) is support for dynamic location based system. For example, Wi-Fi access point could be opened right on the mobile. In this case network infrastructure elements could move. And data linked to them should move too. It is a practical example for dynamic location based systems.

Some of the systems (e.g., [5]) can use a formula that approximates the distance to an access point as a function of signal strength. Using an optimization technique, such systems compute location to an accuracy of about 10 meters using signal strengths from multiple access points. Our own Spot Expert system (SpotEx) [2] is a good example of the empirical approach to Wi-Fi proximity. It is based on the logical rules (productions) and does not require calibration phase. It is the main goal for our systems – use network proximity without the preliminary scene preparation.

The rest of the paper is organized as follows. Section II describes our SpotEx approach. In Section III we describe context-aware check-ins. In Section IV we discuss the usage of Wi-Fi proximity in trajectories calculations.

## II. SPOT EXPERT

SpotEx service presents a new model for context-aware data retrieval. Our model uses only a small part from algorithm of any Wi-Fi based positioning system - the detection of Wi-Fi networks. Due to the local nature of Wi-Fi networks, this detection already provides some data about location, namely information about proximity. As the second step, SpotEx introduces an external database with some rules (productions or if-then operators) related to the Wi-Fi access points. Typical examples of conditions in our rules are: AP with SSID Café is visible for mobile device; RSSI (signal strength) is within the given interval, etc. Based on such conclusions, we can deliver (make visible) user-defined messages to mobile terminals. In other words, the visibility of the content depends on the network context (Wi-Fi network environment).

The SpotEx model does not require a calibration phase. It is based on the ideas of proximity. Proximity based rules replace location information, where Wi-Fi hot spots work as presence sensors. The SpotEx approach does not require mobile users to be connected to the detected networks; it uses only broadcast SSID for networks and any other public information.

Technically, SpotEx contains the following components:

- Server side infrastructure including a database (store) with productions (rules), rules engine and rules editor. The rules editor is a web application (also supporting mobile web) that work with the rules database. This rules engine is responsible for runtime calculations. Note, that the database is currently located outside the mobile device but it could also be positioned right on the device. For example, Wi-Fi Direct based deployment could work this way.

 - The mobile application is responsible for getting context information, matching it against the database with rules and visualizing the output.

SpotEx could be deployed on any existing Wi-Fi network (as well as networks especially created for this service) without any changes in the infrastructure. Rules can easily be defined for the network. Each rule is a logical production (if-then operator). The conditional part includes predicates that work with the following data, measured by the mobile application:

- Wi-Fi network (SSID, mac-address)
- RSSI (signal strength - optionally)
- Time of the day (optionally)
- Client's ID (MAC-address)

Here is a typical rule:

*IF  IS_VISIBLE('mycafe') AND FIRST_VISIT() THEN*

*{ present the coupon info }*

Conclusions for rules (previously termed messages) here are simply texts that should be opened (delivered) to the end-user's mobile terminal as soon as the appropriate rule is fired. For example, as soon as one of the networks is detected via the mobile application. The term "delivered" here is a synonym for "being available for reading/downloading". For end-users the whole process looks like automatic (and anonymous) check-in. It is just an analogue for check-in processing in Foursquare. Mobile users manually post own location to Foursquare and get back some business-related information (badge). In SpotEx the system automatically checks rules against the exiting network context and get back some business-related information (badge).

Existing use cases target proximity marketing at the first hand. One shop can deliver proximity marketing materials right to mobile terminals as soon as the user is near some selected access point. Rather than checking-in (manually or via an API) at a particular place and get back some information on special offers (similar to Foursquare, Facebook Places, etc.), mobile subscribers can collect deals information automatically with SpotEx. We should note, that users do not need to publish own information (location) back to the social networks. Actually, the whole process is anonymous. You do not need to disclose your identity just for looking some local offerings. Figure 1 demonstrates the typical console in SpotEx:

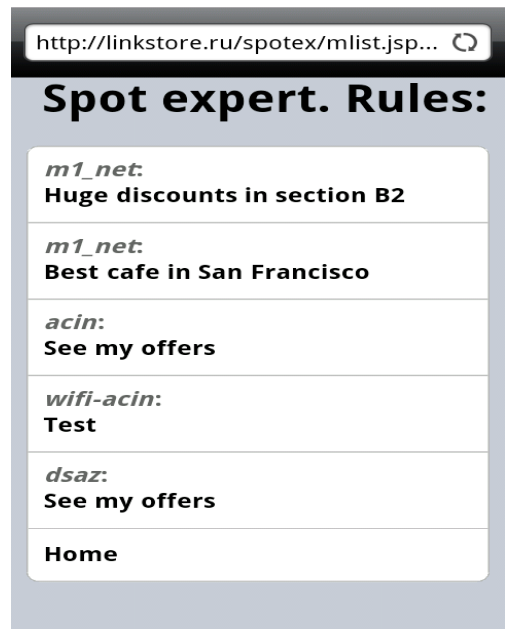

Figure 1.   SpotEx console [2]

Rules in SpoteEx present the standard production rule based system. It means, that the processing could be performed via well know Rete algorithm [6].

The prospect areas, in our opinion, are information systems for campuses and hyper local news delivery in Smart City projects. Rules could be easily linked to the available public networks [7].

### III.    CONTEXT-AWARE CHECKINS

In general, checking rules (conditions) in SpotEx looks like a special check-in service for social networks. What is a typical check-in record in existing social networks? It is some message (post, status) linked to some location (place). What are the reasons for members in social networks use such special kind of messages (posts)? Sometimes it could be stimulated by the business. Practically, user posts advertising for the business in exchange for some benefits. Sometimes it could be used for social connections. Checkin records let other know where I am and see where my friends are.

But the key point is the conclusion, that check-in is just a special kind of record in the social network. It could be customized of course. Business can create own forms for check-in records [8]. QRpon describes a new model for check-in process: customized (or business-oriented) check-ins. This approach introduces a new mobile service that lets any business publish customized records (statuses) in social networks (Facebook in the current release) in exchange for some benefits (discounts, gifts, coupons) provided for the customers. For the business, this service introduces a new way for advertising in the social networks. For the consumers, this service introduces a way for exchanging access to the own social graph for some benefits (e.g., gifts, discounts, coupons). Technically, it is implemented as a production based expert system. But check-ins are still some special posts in the social networks. In other words, they are always part of the social stream.

What if we create a new type of check-in records and separate them from rest of stream? It means, that we will provide a separate database with a list of accounts from the social network. This database specifies accounts concentrated (at this moment only) nearby some place. It is a temporal database, check-in records could be changed constantly and database does not contain the social stream itself. It contains IDs (e.g. nick names) for accounts only.

The second key moment is a new definition for places. Traditionally, for social networks it is a pair from geo coordinates (latitude, longitude) and some description. This definition already can create some problems for indoor systems. For example, many physically different places within the big mall will be on the same geo position, etc. Our idea is to remove our "new" places from the social networks. We will describe our "places" separately and define them via proximity (network proximity in our case) attributes rather than via geo coordinates.

Proximity based definition means, that each place should be defined via some metric that we introduce for our network. Let us talk about Wi-Fi. As a base for metric we can use a list of visible SSID for networks and RSSI (signal level) for the each device. In practice, it is so called Wi-Fi fingerprint.

If two users (mobile phones) have at least one common visible access point – they are in the proximity. And obviously, the more similar (close each other) the visible network environments are than the level of proximity is bigger.

A classical approach to Wi-Fi fingerprinting [9] involves RSSI (signal strength). The basic principles are transparent. At a given point, a mobile application may hear ("see") different access points with certain signal strengths. This set of access points and their associated signal strengths represents a label ("fingerprint") that is unique to that position. The metric that could be used for comparing various fingerprints is k-nearest-neighbors in signal space. It means, that two compared fingerprints should have the same set of visible access points and they could be compared by calculating the Euclidian distance for signal strengths.

Fingerprinting is based on the assumption that the Wi-Fi devices used for training and positioning measure signal strengths in the same way. Actually, it is not so (due to differences caused by manufacturing variations, antennas, orientation, batteries, etc.). To account for this, we can use a variation of fingerprinting called ranking. Instead of comparing absolute signal strengths, this method compares lists of access points sorted by signal strength. For example, if the positioning scan discovered $(SS_A; SS_B; SS_C) = (-20; -90; -40)$, then we replace this set of signal strengths by their relative ranking, that is, $(R_A; R_B; R_C) = (1; 3; 2)$ [9]. As the next step, we can compare the relative rankings by using the Spearman rank-order correlation coefficient [10].

We can use signal strength features for distance estimation in terms of the Euclidean distance in signal strength space and the Tanimoto coefficient [11].

As a prerequisite we compute the vector of average signal strength per access point $S'_x$ from the list of signal strength vectors $S_x$. In the Euclidean version the distances are defined as follows for each pair of average signal strength vectors $S'_a$, $S'_b$, with entries for non-measurable access points in either vector set to -100dBm:

$$d_{a,b} = \|S'_a - S'_b\|$$

For the Tanimoto coefficient version the distance is computed as follows so the value increases as the vectors becomes more dissimilar:

$$d_{a,b} = 1 - (S'_a \cdot S'_b)/(\|S'_a\|^2 + \|S'_b\|^2 - S'_a \cdot S'_b)$$

Now, suppose we have a mobile application that lets users confirm their social network identity and link that identity with wireless networks info. This wireless info is simply the same Wi-Fi fingerprint we've described above for SpotEx. Our application will form (fill) our external temporal database that contains social network identity confirmation and an appropriate network info (Wi-Fi fingerprint). Running this application is simply an act of performing new form of check-in.

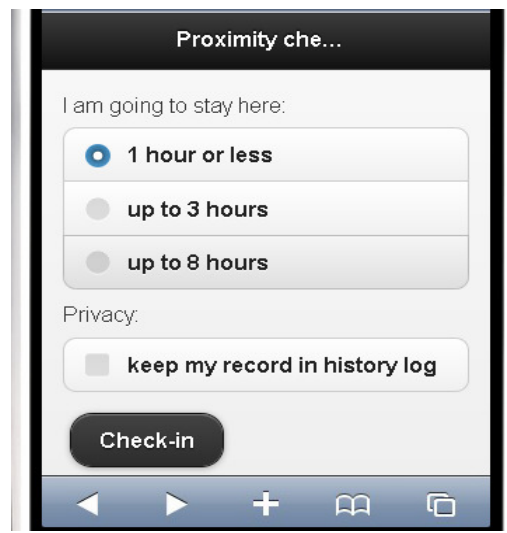

Figure 2. In-proximity check-in

This new check-in is an "external" entity for the social network. Our application does not post data back to the social network. It keeps data outside (in the own database). So, in the terms of privacy, this check-in does not affect (does not touch, actually) account's settings in the social network. In the same time, the real IDs from the existing social networks let us propose discovery mechanism for the social graph. It simply answers to the question: how to discover new members outside of your social graph? The build-in check-ins discovers places for our existing friends. External check-ins database discovers potential new friends.

What are the reasons for users to perform new check-ins? Actually, they are the same as for "old" check-ins. Business entity can use that information for statistics and deliver some benefits in exchange for "check-in". Of course, it could be used for creating the connections too. Our new check-ins let other know where the author is, as well as see who else is there. We can follow to the schema presented in [8], for example. This service at the moment of ID confirmation (it does not post data back to social network, but ID confirmation is required) has access to the user's social graph. It means, particularly, that we can program output (our confirmation page) depends on the social graph size, for example. In other words, the confirmation screen generation might be actually some production (rule based) system. It could be a set of rules like this:

*IF* (some condition) *THEN* (some conclusion)

Our conditional part includes a set of logical operations against user's social graph data and conclusion is our output (coupon, gift confirmation, etc.). So we are going to say here, that this external check-in system could be actually some sort of expert systems (production based) that generates conclusions (badges) by the social graph defined conditions.

We can provide an interesting set of new services with this external check-ins database. At the first hand, we can list other people at any particular location. Actually, it is always a list of people at "this" location only. It follows from the fact, that our proximity based system does not provide a list of locations in the traditional form. Each our "location" is described via Wi-Fi sensors (via Wi-Fi fingerprints snapshot). Obviously, all the attributes are dynamical. So, as soon as the check-in is performed, user can see nearby check-ins only. In this approach user is simply unaware about other "places" unless he moves and performs a new check-in.

At the second, we can show (search) social streams nearby. Via public API, we can read data feeds for users (if it is possible, of course, and an appropriate data stream is not protected). This system can keep the full respect to the existing privacy settings in social networks.

IV. WI-FI PROXIMITY AND TRAJECTORIES

SpotEx service can collect Wi-Fi snapshot information during the execution. It is not the above mentioned preliminary Wi-Fi scene preparation. We talk here about real-time data. Besides checking the rules, our discovery process can simply collect wireless network environment snapshots (Wi-Fi fingerprints, actually). SpotEx application collects them from the moment user started the application. Note, that because there is no authorization in SpotEx, this information is completely anonymous.

Obviously, this collected information could be used for data discovery too. Of course, we can investigate historical logs too, but it is separate task. It is similar to the models presented in Reality Mining projects [12], for example. In the classical paper, authors perform cluster analysis for the previously collected data. A Hidden Markov Model conditioned on both the hour of day as well as weekday or weekend provided data separation for behavior patterns like "home", "office", etc.

In general, collecting this information is yet another example of using mobile phone as a sensor. And of course, SpotEx is not the only approach uses phone as a sensor concept. As per this task, we've tested the ability to implement our approach with the project Funf [13], for example. Funf Probes are the basic collection data objects used by the Funf framework. Each probe is responsible for collecting a specific type of information. These include data collected by on-phone sensors, like accelerometer or GPS location scans, etc. Actually, in Funf many other types of data (context info) can be collected through the phone. In other words, Funf is a rich data logger. For our network proximity based project we need only a small part of this logger. We are using the part that collects information about wireless environment.

The first idea has been implemented with proximity logs targets convoys detection. As soon as the proximity information replaces the location data (and it is especially important for the indoor applications), we should be able to replace the traditional location-based calculations with proximity based measurements. One of such interesting calculations is the detection of convoys. Convoy is a group of objects that travel together for more than some minimum duration of time. More probably, that the original task for discovery of convoys (groups of objects with coherent trajectory patters) was oriented to the military applications. As per nowadays research papers, a number of applications may be envisioned. The discovery of common routes among citizens may be used for the scheduling of public transport. The discovery of convoys for trucks may be used for throughput planning, etc. Figure 3 illustrates the convoys:

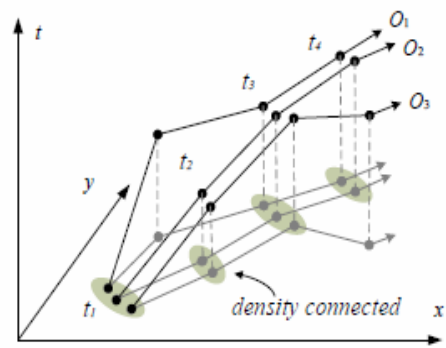

Figure 3. Convoys [14]

Detection of convoys on the base of Wi-Fi proximity logs let us detect in SpotEx group of people. For example, in proximity marketing, some shop (café, etc.) can present a special program for the grouped visitors.

Shortly, convoy is a group of moving objects, which are in density connection the consecutive time points. Objects are density-connected, if a sequence of objects exists that connects the two objects and the distance between consecutive objects does not exceed the given value [14,15].

The next often used in this context terminology is moving cluster (or cluster of moving objects) [16]. The moving cluster exists, if a shared set of objects exists across some finite time, but objects may leave and join a cluster during the cluster's life time. So, the semantic is different and moving clusters do not necessarily qualify as convoys (in the pure terms). Another interesting term in this space is flock. Flock is a set of objects that travel within a range while keeping the same motion. Considering SpotEx related task, we will use our own definition for moving clusters [17].

Let us start from the collected data. Wi-Fi environment (collected by SpotEx or Funf) is a time stamped list of records. Each environment's record is a vector of triples. Each triple describes one Wi-Fi network:

Network ID (SSID)

MAC-address

signal strength (RSSI)

and the whole environment could be described as a vector of triples:

$E = \{T_1, T_2, ..., T_n\}$

Our fingerprint is just a time stamped environment: $[t_i, E_i]$.

Two networks measurements are comparable in this paper (it is the simplest metric) if they have at least one common access point with difference in signal strengths less than the given threshold. Technically, with this proximity versus location model, we are going to add to SpotEx's list of predicates a new logical function:

IN_GROUP_OF (n, t)

Here $n$ presents some positive integer value and $t$ describes a time (e.g., seconds). This function returns Boolean value *true* if mobile user traveled in the group of at least n people during at least t seconds. It is, by the SpotEx vision, of course. In other words, all those *n* people should be presented via own records in the proximity log. We think, that such a function (actually, it is a new qualification for our context) could be useful in proximity marketing tasks, for data discovery in Smart City projects, etc.

Initial parameters for our algorithm are:

$\Delta$ - time threshold, $\Omega$ - RSSI threshold, $E$ - an original network environment, $T_0$ – an original current time, $T_{max}$ – argument for function

1. Initialize new candidate set $R_1$
2. Collect measurements within the time $T_0$-$\Delta$ $\rightarrow$ $R_1$;
3. **If** $R_1$ is empty **then** output *false;*
4. Remove from $R_1$ all measurements that are not comparable with $E$;
5. **If** $R_1$ is empty **then** output *false;*
6. **Set** $t = T_0$;
7. **While** $t > T_0$-$T_{max}$

8. Find the previous measurement for the current user. Update current settings $\rightarrow \{t, E\}$;
9. For the each measurement in $R_1$ find proximity data within $t \pm \Delta$ (update measurements with new data);
10. Remove from $R_1$ elements without new data (not updated elements) ;
11. Remove from $R_1$ elements that are not comparable with $E$;
12. **If** $R_1$ is empty **then** break;
13. **End while**

The finally, $R_1$ presents the group we are looking for. Depends on the size of this array, we can calculate our function IN_GROUP_OF( ).

V. CONCLUSION

This paper describes several mobile services based on the common approach to Wi-Fi proximity. In general, our approach uses smart-phone as a proximity sensor. Described services can use existing as well as the especially created networks nodes as presence triggers for delivering and discovering new content for mobile subscribers. Proposed services could be used for proximity marketing and delivering commercial information (deals, discounts, coupons) in malls, for access to hyper-local news data and for data discovery in Smart City projects.